\documentstyle[12pt,epsf,epsfig]{article}
\newcommand{\btab}{\begin{tabbing}}
\newcommand{\etab}{\end{tabbing}}

\newcommand{\beqn}{\begin{equation}}
\newcommand{\eeqn}{\end{equation}}
\newcommand{\barr}[1]{\begin{array}{#1}}
\newcommand{\earr}{\end{array}}
\newcommand{\beqna}{\begin{eqnarray}}
\newcommand{\eeqna}{\end{eqnarray}}
\newcommand{\btablec}{\begin{table} \begin{center}}
\newcommand{\etablec}{\end{center} \end{table}}

\newcommand{\gapproxeq}{\lower.7ex\hbox{$\;\stackrel{\textstyle>}
{\sim}\;$}}
\newcommand{\tenbar}{{\overline {10}}}
\newcommand{\plabel}[1]{\label{#1}}
\newcommand{\pbibitem}[1]{\bibitem{#1}}
\def\question#1{}
\input epsf

\begin{document}
\title{
\begin{flushright} 
\vskip -1.0 cm
%\small{hep-ph/0307019} %\\ 
%\small{LA-UR-03-XXXX}  
\end{flushright} 
\vspace{0.6cm}  
\Large\bf Interpretation of the $\Theta^{+}$ as an isotensor
pentaquark with weakly decaying partners}
\author{Simon Capstick\thanks{\small \em E-mail: capstick@csit.fsu.edu} \\
{\small \em Department of Physics, Florida State University,
Tallahassee, FL 32306-4350, U.S.A.}\\
Philip R. Page\thanks{\small \em E-mail: prp@lanl.gov} \\
{\small \em Theoretical Division, MS B283, Los Alamos
National Laboratory,}\\ 
{\small \em  Los Alamos, NM 87545, U.S.A.} \\
Winston Roberts\thanks{\small \em E-mail: roberts@qcd.physics.odu.edu} \\
{\small \em Department of Physics, Old Dominion University, Norfolk, VA
23529, U.S.A.} \\
{\small \em Theory Group, Thomas Jefferson National Accelerator Facility,}\\
{\small \em 12000 Jefferson Avenue, Newport News, VA 23606, U.S.A.} }
%\date{September 1998}
\date{}
\maketitle
\begin{abstract}
The $\Theta^+$(1540), recently observed at LEPS, DIANA and CLAS, is
hypothesized to be an isotensor resonance. This implies the existence
of a multiplet where the $\Theta^{++}$, $\Theta^{+}$ and $\Theta^{0}$
have isospin-violating strong decays, and the $\Theta^{+++}$ and
$\Theta^{-}$ have weak decays and so are long-lived. Production
mechanisms for the weakly-decaying states are discussed. The $J^P$
assignment of the $\Theta$ is most likely $1/2^-$ or $3/2^-$.
\end{abstract}
\bigskip

PACS number(s): 11.30.Hv,\hspace{.2cm} 12.15.-y,\hspace{.2cm}
12.39.Mk,\hspace{.2cm}14.80.-j


\section{The $\Theta^+$(1540) as an isotensor pentaquark}

Recently the LEPS Collaboration at SPring-8 reported the $4.6$ $\sigma$
discovery of a new resonance, $\Theta^+$, in the reaction
$\gamma\,^{12}{\rm C}\rightarrow$ C$' K^- \Theta^+ \rightarrow$
C$^\prime K^- K^+ n$, with a mass of $1.54\pm 0.01$ GeV and a width of
less than $25$ MeV~\cite{leps}.  Subsequently, the DIANA Collaboration
at ITEP reported a $4.4$ $\sigma$ discovery of the {\it same}
resonance in $K^+$ Xe $\rightarrow \Theta^+$ Xe$'\rightarrow K^0 p$
Xe$'$ with mass $1539\pm 2$ MeV and width less than $9$
MeV~\cite{diana}.  Preliminary results from an experiment by the CLAS
Collaboration in Hall B at Jefferson Lab confirm the existence of a
narrow $\Theta^+$ in the reactions $\gamma d \to \Theta^+ K^- p \to
K^+ (n) K^- p$ at a mass of $1542 \pm 5$ MeV and width less than $22$
MeV, and $\gamma p \rightarrow \Theta^+ \pi^+ K^-
\to K^+(n)\pi^+K^-$, at around the same mass~\cite{stepanyan}.

The $\Theta^+$ is interpreted as a state containing a dominant
pentaquark Fock-state component $uudd\bar{s}$ decaying to $K^+ n$, in
contrast to its interpretation as a chiral soliton~\cite{chiral,DPP}.
The pentaquark should have isospin $I=0$, $1$ or $2$. Here we
hypothesize a certain isospin assignment for this state based on two
precepts:

$\bullet$ The $\Theta^+$ seen experimentally exists, and is resonant.

$\bullet$ A pentaquark 110 MeV above threshold should have a decay
width of the order of $500$ MeV unless its decays are suppressed by
phase space, symmetry, or special dynamics.

As will be clear from the discussion of the second precept below, the
expected width for the decay $\Theta^+ \rightarrow K^+ n$ is greater
than the experimental widths unless the decay proceeds in F-wave or
higher, i.e. the total angular momentum $J$ of $\Theta^+$ is $\geq
5/2$ for a parity $P=+$ $\Theta^+$, or $J\geq 7/2$ for $P=-$. We
consider such a high $J$ assignment for the $\Theta^+$ unlikely for
such a light resonance (see the discussion of the $J^P$ assignments of
the $\Theta^+$ below). Hence the two precepts indicate that the decay
of the $\Theta^+$ is suppressed either by special dynamics unknown to
us, or by symmetry. It is hypothesized that the latter possibility
ensues, and specifically that $\Theta^+$ is $I=2$. For this isospin,
the decay $\Theta^+ \rightarrow K^+ n$ is isospin-symmetry violating.
This is not the case if the $\Theta^+$ is isoscalar or isovector.  It
is shown below that isospin symmetry violating decay widths are
typically $0.1\%$ of isospin conserving widths.

The reasons for the second precept are as follows. Any state with the
structure $qqqq\bar{q}$, for any flavors, has as a possible color
configuration a set of three quarks in a colorless (baryon) state,
plus a quark and anti-quark in a colorless (meson) state. This means
that by a simple rearrangement of the color configuration, the state
can ``fall apart'' into a baryon and meson, with only weak forces
between the two colorless hadrons. Multi-quark states which are above
threshold for such fall-apart decays can be expected to be immeasurably
broad, as no such states have been observed to date. The WA102
collaboration~\cite{Barberis} sees roughly 15,000 events in
$f_2(1270)\pi\pi$ that are well described by a Breit-Wigner at 1950
MeV with a width of 450 MeV, associated with the state
$f_2(1950)$. This is likely the broadest well established
resonance. This would suggest that multi-quark states that fall-apart
should have decay widths of at least 500 MeV.

The largest phase space for the fall apart decay of an isoscalar or
isovector $\Theta^+\to nK^+$ will be when the $\Theta^+$ has
$J^P=1/2^-$, and so decays in an $S$-wave. This is the lowest angular
momentum possible for the $\Theta$ if its four quarks and anti-quark
are in a spatial ground state. If the width of such a state is 500
MeV, then other spin-parity assignments for the $\Theta^+$ will allow
decay widths suppressed by phase-space factors
$(p/\beta)^{2L}/(2J+1)$, where $\beta\simeq 400$ MeV is a typical
energy scale associated with the size of hadrons, and $p=270$ MeV is
the center of mass decay momentum.

If the $\Theta^+$ is isoscalar or isovector with spin-parity $3/2^-$
or $5/2^-$, also possible in a spatial ground state, the decay is
$D$-wave, with a width significantly larger than the observed
limit\footnote{The state $\Lambda(1520)$ decays to $\bar{K}N$ in
$D$-wave, has a mass similar to the $\Theta^+$, but has a small width
of $15.6\pm 1.0$ MeV. However, its decay requires the creation of a
light $q\bar{q}$ pair through a non-trivial decay operator. It is,
therefore, necessarily narrower than a similar state with a fall-apart
decay, where the final state configuration exists in the initial state
and so the decay operator is trivial (unity).}. If the $\Theta^+$ has
spin-parity $1/2^+$ or $3/2^+$, which require one unit of orbital
angular momentum, the decay is $P$-wave with larger widths. If the
$\Theta^+$ has $J^P=5/2^+$, $7/2^-$,... then it decays in $F$-wave or
higher, and it is possible that is an isoscalar or isovector state and
phase space suppression accounts for the small observed
width. However, it is unlikely that such a state would be a ground
state, as the addition of orbital angular momentum will significantly
increase its energy.

In what follows, an estimate is made of the width of the
isotensor $\Theta^+\to nK^+$ decay.
Isospin-violating strong decays in mesons have widths which are
typically fractions of MeV. The $P$-wave decay $\omega\to \pi^+\pi^-$
is a $G$-parity and so isospin violating strong decay, with a partial
width of 0.14 MeV, which is roughly $0.1\%$ of the isospin conserving
$\rho\rightarrow\pi\pi$ width.

An upper bound on the width of an isotensor $\Theta^+$ can be
estimated as follows. The $m_d-m_u$ mass difference and
electromagnetic interactions cause the wave functions of the
$\Theta^+$ and final-state nucleon to contain small isospin impurities
\begin{eqnarray*}
\vert \Theta^+\rangle&=&\vert I=2,I_z=0\rangle+\alpha \vert 1,0\rangle
+\beta \vert 0,0\rangle\; ,\\
\vert n\rangle&=&\left\vert 1/2,-1/2\right\rangle+\gamma \left\vert 
3/2,-1/2\right\rangle\; .
\end{eqnarray*}
Assuming a strong decay operator ${\cal O}$ which conserves isospin,
the isospin violating $\Theta^+\to nK^+$ decay width will be
\[
\left\langle \Theta^+\vert {\cal O} \vert n K^+\right\rangle
\simeq \gamma \left\langle 2\vert {\cal O} \vert 3/2,1/2\right\rangle
+\alpha \left\langle 1\vert {\cal O} \vert 1/2,1/2\right\rangle
+\beta \left\langle 0\vert {\cal O} \vert 1/2,1/2\right\rangle \; ,
\]
ignoring terms second order in the small coefficients and suppressing
$I_z$ values. Isospin-violating effects in the nucleon wave function
are known to be very small~\cite{Isgur,Miller}, so $\gamma$ is
negligible. A larger mixing occurs in the $\Lambda^0$--$\Sigma^0$
system, where the physical $\Lambda$ is a mixture
$\Lambda^0_8+\epsilon\,\Sigma^0_8$ of states of definite isospin with
a mixing coefficient~\cite{Karl} of $\epsilon=-0.015$, and similarly
for the physical $\Sigma^0$. Assuming that $\alpha$ and $\beta$ are
both of this size, and that the isospin-conserving decays widths (of
our state well above its fall-apart decay threshold) are all of the
order of $500$ MeV, a rough upper bound for the $\Theta^+$ width is
$4\,(0.015)^2\,(500\ {\rm MeV})=0.45\ {\rm MeV}$. This is of the same
order of magnitude as that of the isospin-violating $\omega\to
\pi^+\pi^-$ decay width, and is much smaller than the experimental
upper bounds on the $\Theta^+$ width.

\section{Prediction of strongly decaying $\Theta^{++}$ and $\Theta^{0}$ 
and\\ weakly decaying $\Theta^{+++}$ and $\Theta^{-}$}

\begin{table}[t]
\begin{center}
\begin{tabular}{|l||l|r|c|}
\hline %------------------------
State & Quarks & $I_z$ & Decay modes \\
\hline \hline %------------------------------
$\Theta^-$ & $dddd\bar{s}$ & $-2$ & \\
$\Theta^0$ & $uddd\bar{s}$ & $-1$ & $n K^0$\\
$\Theta^+$ & $uudd\bar{s}$ & $0$ & $n K^+$, $pK^0$\\
$\Theta^{++}$ & $uuud\bar{s}$ & $1$ & $p K^+$\\
$\Theta^{+++}$ & $uuuu\bar{s}$ & $2$ & \\
\hline %------------------------------
\end{tabular}
\caption{\plabel{tab1} Quark content, $I_z$, and strong decay modes 
of $\Theta$ states.}
\end{center}
\end{table}

If the $\Theta^+$ is an $I=2$ pentaquark, it is the isospin projection
$I_z=0$ member of the multiplet depicted in Table~\ref{tab1}. If
discovered, the $\Theta^{+++}$ would be the first triply charged
hadron. Typical splittings in $I=\frac{1}{2}, 1$ or $\frac{3}{2}$
multiplets are $n-p=1.3$ MeV, $\pi^{\pm}-\pi^0=4.6$ MeV and
$\Delta^0-\Delta^{++}=2.3$-$2.7$ MeV respectively~\cite{pdg}. From
model expectations and data~\cite{CW-Delta}, the $\Delta^- -
\Delta^{++}$ splitting is likely close to 4.5 MeV. Based on this, the
largest mass splitting in the $\Theta$ multiplet is expected to be
less than $10$ MeV.  The relevant threshold for strong decay of the
$\Theta$ is $N K$. Since the $\Theta^+$ is below $N K \pi$ threshold
by about 30 MeV, its isospin partners are also\footnote{The lowest
threshold is $p\pi^0K^\pm$ at 1567 MeV. Only the $p\pi^0K^+$ decay can
come from a pentaquark with the structure of the $\Theta^+$.}.  This
precludes the strong decays $\Theta\rightarrow N K \pi$, specifically
$\Theta^-\not\to n \pi^- K^0$ and $\Theta^{+++}\not\to p \pi^+ K^+$,
and so these states must\footnote{The electromagnetic decay $\Theta
\rightarrow$X$\gamma$ is kinematically forbidden if there is no
pentaquark X with the charge of the $\Theta$, but at a lower mass.}
decay weakly as indicated in Table~\ref{tab2}.  A multiplet where the
central members ($\Theta^0,\Theta^+$ and $\Theta^{++}$) decay
strongly, while the outlying members ($\Theta^{+++}$ and $\Theta^-$)
decay weakly, has no analogue for known mesons and baryons.

\begin{table}[t]
\begin{center}
\begin{tabular}{|l||l|c|c||l|l|c|c|}
\hline %------------------------
State & Decay mode & & Pairs & State & Decay mode & & Pairs \\
\hline \hline %------------------------------
$\Theta^{+++}$ & $p\pi^+     l^+\nu_l$      &A & 1& $\Theta^{-}  $ & $n\pi^-$                   &E & 0\\
               & $p\pi^+     \pi^+$         &A & 1&                & $n\pi^-     \pi^0$         &E & 1\\
               & $p\pi^+     \pi^0l^+\nu_l$ &P & 1&                & $n\pi^-     \pi^-l^+\nu_l$ &P & 1\\
               & $\Delta^{++}l^+\nu_l$      &A & 0&                & $\Delta^-\pi^0$            &E & 0\\
               & $\Delta^{++}\pi^+$         &A & 0&                & $\Delta^{- }\pi^-l^+\nu_l$ &P & 0\\
               & $\Delta^{++}\pi^0l^+\nu_l$ &P & 0&                & $\Delta^-f_0(600)$         &E & 0\\ 
\hline %------------------------------
\end{tabular}
\caption{\plabel{tab2} Semi-leptonic and non-leptonic weak decay modes
of the $\Theta^{+++}$ and $\Theta^{-}$ in order of increasing phase
space. Here $l=e,\mu$.  All decay modes listed are Cabibbo
suppressed. Also indicated is whether the decay proceeds by single $W$
annihilation (A), exchange (E), or production (P), as well as the
number of $q\bar{q}$ pairs that are created by the strong interaction.
Decays involving final states $\Delta \pi\pi (l^+\nu_l)$, $N\pi\pi\pi
(l^+\nu_l)$ and $N\pi\pi\pi\pi (l^+\nu_l)$ are not listed. Only
two-body decays involving the $f_0(600)$ are indicated. Genuine
three-body decays, e.g. $p\pi^+\pi^+$, are distinguished from those
that proceed through an intermediate resonance,
e.g. $\Delta^{++}\pi^+\rightarrow p\pi^+\pi^+$, even though they lead
to the same final state.  Because the $\Delta$ is broad, a final state
$\Delta X l^+\nu_l$ would be experimentally indistinguishable from
$N\pi X l^+\nu_l$, if $\nu_l$ is undetected.}
\end{center}
\end{table}

\section{Explanation of CLAS and DIANA data}

The isotensor $\Theta^+$ is produced via isospin conserving 
reactions of the form $\gamma N\rightarrow \Theta^+ K$ or
$\gamma d \rightarrow \Theta^+K^- p$, through the
isovector component of the photon, at LEPS and 
CLAS\footnote{The $d$ has $I=0$.}.

In the CLAS collaboration's photo-production experiment with a
deuteron target, the $\Theta^+$ is reconstructed from the invariant
mass spectrum of $K^+ n$ in the reaction $\gamma d \to p K^- K^+
(n)$. The four momenta of all reaction products are specified, as the
three charged particles in the final state are detected, and the
neutron is then identified by missing mass. Note that a detailed study
of this reaction was unable to explain the events interpreted as the
$\Theta^+$ by non-resonant rescattering
processes~\cite{meyer}. Similarly, in the photo-production experiment
with a proton target, the neutron in $\gamma p \to \pi^+K^-K^+(n)$ is
identified by missing mass after detection of the charged particles.

In the deuteron target experiment, it should be possible to produce
the isospin partner $\Theta^{++}$ of the $\Theta^{+}$ by the reaction
$\gamma d\to \Theta^{++} K^- n \to pK^+ K^-(n)$, and detect it by
examining the invariant mass of the $K^+p$ system~\cite{Gao}. It may
be the case that production of the $\Theta^{++}$ is suppressed
relative to that of the $\Theta^{+}$. If, as is common in kaon
photo-production experiments at these photon energies, the reactions
$\gamma n \to K^-\Theta^+$ and $\gamma p\to K^- \Theta^{++}$ result in
forward peaked $K^-$ distributions, these negatively charged particles
will be bent into the beam direction by the magnetic field and will go
largely undetected, unless they scatter off the spectator
nucleon. Scattering cross sections for $K^- n$ are considerably
smaller than those of $K^- p$ where they are measured, at kaon beam
energies of 600 MeV or higher~\cite{pdg}.  So if the $K^-$ particles
are forward peaked and the $K^- n$ cross section remains small down to
low energies, it is likely that the $\Theta^{++}$ will remain
undetected without a larger data sample.

Under ideal circumstances a formation reaction of the form $K^+ n
\rightarrow \Theta^+\rightarrow K^0 p$, with a $K^+$ momentum of $440$
MeV incident on a Xe target at rest and with the Xe$'$ at rest in its
ground state (quasi-free production), could account for the ITEP data.
However, such a formation process is isospin violating for an
isotensor $\Theta^+$. It is hence predicted that ITEP should not see
the $\Theta^+$ through this formation process.  In the ITEP experiment
the kinematics are not completely reconstructed for the reaction $K^+$
Xe $\rightarrow K^0 p$ Xe$'$, as the energy and momentum of the Xe$'$
is unknown, and so it is not possible to know whether this quasi-free
production process happens.  If ITEP is indeed seeing the isotensor
$\Theta^+$, this is expected to be through isospin allowed $K^+
[pn]_{I=1}\to \Theta^+[p]$ or $K^+ [nn]_{I=1}\to \Theta^+[n]$
processes.

\section{Mechanisms for the production of $\Theta^{+++}$ and
$\Theta^{-}$}

Here we focus on likely production mechanisms for the weakly decaying
pentaquark states $\Theta$, as opposed to anti-pentaquark states
$\bar{\Theta}$, because of the availability of nuclear targets. The
figure of merit is whether the production cross-section for $\Theta$
times the branching ratio of $\Theta$ into its final state is
substantial.

On face value the most promising process for the production of the
$\Theta^{+++}$ is $pp\rightarrow \Theta^{+++}\Sigma^-$, which involves
one $s\bar{s}$ pair creation. The search may be feasible at ITEP, JHF
or COSY. The reaction $pp\rightarrow \Theta^{+}\Sigma^+$ was
previously suggested~\cite{polyakov} in order to search for the
$\Theta^{+}$. Another promising process is
$K^+p\rightarrow\Theta^{+++}\pi^-$, involving the creation of one
light quark pair. A search at JHF and ITEP is feasible.  In these
experiments the weakly-decaying $\Theta^{+++}$ could be either
long-lived and produce a track in the detector, or relatively
short-lived, and so should be observed in $p\pi^+\pi^+$ (see
Table~\ref{tab2}).

Production of the $\Theta^{-}$ in $pp$ or $K^+p$ collision would
require more simultaneous quark pair creations, and so is strongly
suppressed.  The most promising production mechanism for $\Theta^{-}$
appears to be $nn\to \Theta^- \Sigma^+$, although this would require a
neutron beam more energetic than the maximum $800$ MeV beam at LANSCE
at LANL.

There is a secondary production mechanism for the $\Theta^{+++}$ and
$\Theta^-$ in photo- or electro-production at JLab, SPring-8, Mainz,
SAPHIR or HERA, following production of a $K^+$ or $K^0$. The
production of the $\Theta^{+++}$ would proceed as follows.  In the
reaction $\gamma d \to \Sigma^-\pi^-\Theta^{+++}$ a $K^+$ can be
produced off the neutron ($\gamma n \to \Sigma^- K^+$), which then
interacts with the proton, $K^+p\to \Theta^{+++}\pi^-$. The final
state $\Sigma^-\pi^-\Theta^{+++}$ consists entirely of weakly decaying
particles.  The process $\gamma d \to \Sigma^-\pi^-\Theta^{+++}$
involves in total two $q\bar{q}$ pair creations, and scattering of the
$K^+$ with the proton, qualitatively similar to the possible
production mechanisms at CLAS discussed in the previous section. The
production mechanism for the $\Theta^-$ is closely related, and occurs
at a similar level in $\gamma d \to \Sigma^+\pi^+\Theta^-$. Here the
effect is to produce a $K^0$ off the proton ($\gamma p \to \Sigma^+
K^0$), which then interacts with the neutron, $K^0 n\to
\Theta^{-}\pi^+$.

\section{Simple pentaquark model and $J^P$ assignments}

It is conceivable to interpret the isotensor $\Theta^+$ as a $\Delta
K$ molecule below threshold. However, this interpretation faces two
problems:

$\bullet$ The $\Delta$ is short-lived ($115-125$ MeV wide~\cite{pdg}), 
making a simple molecular picture unlikely.

$\bullet$ The $\Theta^+$ is $\sim 190$ MeV below the $\Delta K$
threshold, which is an atypically large binding energy compared to
molecular candidates like the $f_0(980)$ and $a_0(980)$, which are
$\sim 10$ MeV below threshold. Furthermore, relating the binding
energy of a molecule $1/E_b \sim 2\mu\, r^2_{\mbox{r.m.s.}}$ to the
root mean square separation between the $\Delta$ and $K$, yields
$r_{\mbox{r.m.s.}}\sim 0.5$ fm. This distance is smaller or similar to
the sizes of the constituents, so that the picture of a molecule built
from undeformed hadrons is not reasonable.

%$\bullet$ The $\Theta^+$ must be deeply bound based on a quark model
%argument: The mass of the nucleon in the non-relativistic quark
%model approximately equals the sum of the mass of its constituents:
%$3\times 330$ MeV. On the contrary, the sum of the constituent masses
%in the $\Theta^+$ is $4\times 330 + 550$ MeV. This is $330$ MeV
%{\it above} the mass of the $\Theta^+$.

The $\Theta^+$ is hence best modeled as a pentaquark, as opposed to a
loosely bound molecular state.  An isotensor pentaquark can only be
constructed when each of the quark-pairs are isovector, which means
that each of the quark-pairs $ij$ and $kl$ must be symmetric under
exchange $i\rightarrow j$, $k\rightarrow l$, so that the flavor wave
function is also totally symmetric.  Assuming that the $\Theta^+$ is
the ground-state pentaquark, and that the ground-state has all the
quarks and antiquark in relative S-waves (which would give the lowest
energy if there is no strong repulsive core), the spatial wave
function is totally symmetric. The Pauli principle and isospin
symmetry require the overall fermion wave function to be totally
antisymmetric under exchange of the four light quarks, so that the
color-quark-spin wave function of the four quarks must be totally
antisymmetric.

This implies that the four quarks must be in an antisymmetric
representation of the color-spin symmetry group SU(6). This
is~\cite{Lipkin} the ${\overline {\bf 15}}$ representation of SU(6),
which is made up of a color ${\overline {\bf 6}}$ with spin zero, and
a color triplet of spin 1. When combined with the color $\overline{\bf
3}$ antiquark, a color singlet pentaquark can only have the four
quarks combined to a color-triplet with spin 1. This restricts the
isotensor pentaquark to $J^P=1/2^-$ or $3/2^-$. These $J^P$ would only
allow $S$- or $D$-wave decays $\Theta^+\to nK^+$, which, when compared
with experiment, indicates that $\Theta^+$ is anomalously narrow. Even
though the observation of $\Theta^+$ in $nK^+$ allows for any $J^P$
assignment, and an angular analysis must yet be performed to determine
the parity of the $\Theta^+$, it is expected that $J^P = 1/2^-$ or
$3/2^-$.

In Ref.~\cite{stancu} the $\Theta^+$ is interpreted as an isoscalar or
isovector pentaquark in a constituent quark model with a flavor-spin
interaction that arises from Goldstone boson exchange between the
quarks. This flavor-spin interaction provides an attractive
interaction between the quarks, which can lower the energy of the
state with the four light quarks in an $P$-wave state below that of
the corresponding $S$-wave state, so that the ground state pentaquark
has positive parity~\cite{stancu2}.

\section{Discussion}

The chiral-soliton model prediction~\cite{DPP} of a narrow isoscalar
state $Z^+$ at 1530 MeV, situated at the top of an SU(3)$_f$
anti-decuplet of $J^P=1/2^+$ states, partially motivated the searches
for the experimental $\Theta^+$ state described above. In
Ref.~\cite{DPP}, the width of the $Z^+$ is predicted to be 15 MeV. The
mass of the $Z^+$ is found by ``anchoring'' the anti-decuplet to a
non-exotic member of the anti-decuplet, which has nucleon flavor. This
is identified with the established state $N(1710)$ seen in $N\pi$
elastic scattering, as well as in other channels. 
If the anti-decuplet $P_{11}$ is chosen to be N(1440),
the $Z^+$ state would be stable against strong decays, 
and if it is chosen to be N(2100) the $Z^+$ would be very broad. 
The $N(1710)$ state is described in the constituent
quark model as a radial excitation of the nucleon, and its strong
decays, photocouplings, and mass~\cite{CIW} are well understood in that model.

The total width of the $N_\tenbar(1710)$ state is predicted in the chiral
soliton model to be 43 MeV, which is below the 50 to 250 MeV range
quoted by the Particle Data Group (PDG)~\cite{pdg}, and significantly below
the PDG's ``estimate'' of 100 MeV. Similarly, the width of the
$\Sigma_\tenbar$, associated with the two-star $P_{11}$ state
$\Sigma(1880)$, is predicted to be about 70 MeV, which is again
smaller than the 80-260 MeV range of widths quoted by the PDG.  In the
chiral soliton model, the widths of all members of the anti-decuplet are
proportional to a calculated constant $G^2_\tenbar$. If this was
adjusted upward to accommodate the PDG estimate of the width of the
$N(1710)$, the predicted width of the $Z^+$ would increase, and would
exceed the bounds set by the experimental data.

The essential mechanism for the narrow width of the $\Theta^+$, which
is well above its strong $n K^+$ and $pK^0$ thresholds, is that its
decay to $n K^+$ or $pK^0$ is isospin violating if it is isotensor.
It is as though the strong decay threshold has been ``raised''. It is
intriguing to note that such raised thresholds will occur for all
isotensor $qqqq\bar{Q}$ states, with $q$ an up or down quark, and $Q$
a heavy quark, since their decay to $N (q\bar{Q})$ also violates
isospin conservation.  A related phenomenon happens for the
$dddQ\bar{u}$ and $uuuQ\bar{d}$ pentaquarks.  The decays
$dddQ\bar{u}\to \Delta^- (Q\bar{u})$ or $\Sigma_Q\pi^-$ are allowed,
while $dddQ\bar{u}\to N (Q\bar{u})$ or $\Lambda_Q\pi$ are not. Also,
the decays $uuuQ\bar{d}\to \Delta^{++} (Q\bar{d})$ or $\Sigma_Q\pi^+$
are allowed, while $uuuQ\bar{d}\to N (Q\bar{d})$ or $\Lambda_Q\pi$ are
not.

\section*{Acknowledgments}

Helpful discussions with T.~Barnes, T.~Goldman, R.~C.~Haight,
K.~Hicks, V.~Koubarovski, H.~J. Lipkin, C.~A.~Meyer, R.~Schumacher,
and S.~Stepanyan are gratefully acknowledged. This research is
supported by the U.S. Department of Energy under contracts
DE-FG02-86ER40273 (SC), W-7405-ENG-36 (PRP), DE-AC05-84ER40150 (WR)
and DE-FG05-94ER40832 (WR).

%\appendix

%\section{Appendix: }

\end{document}